\documentstyle[epsfig,aps]{revtex}     

\begin{document}

\title{\vglue -0.5cm
 \hfill{\small IFUSP-DFN/98-008} \\
\vglue .5cm
The Network Solution For Electron Identification in a Wide Momentum 
             Region with a TRD}

\author{N.\ Kuropatkin{\thanks{On leave of absence from Petersburg Nuclear 
Physics Institute,Gatchina, Russia.}$^,$\thanks{E-mail:~kuropat@charme.if.usp.br.}}  and R.\ Zukanovich Funchal\thanks{E-mail: zukanov@charme.if.usp.br.}}

\address{ Instituto de F\'\i sica da Universidade de S\~ao Paulo\\
05389-970 C.P. 66318 -- S\~ao Paulo, SP\\
Brazil}

\maketitle

\begin{center}
{\em (To be published in the Proceedings of the International Conference 
on Computing in High Energy Physics 98 )}
\end{center}

\hfuzz=25pt

\begin{abstract}
  A Feed Forward Error Back Propagation Artificial Neural Network~(ANN) 
 algorithm is developed for electron/positron identification in a wide
  momentum region (10 - 300 GeV/c). The method was proposed for the Transition
  Radiation Detector of the E781 experiment at Fermilab. The package consists 
   of two parts:

    $\bullet$ the program for the ANN training;

    $\bullet$ the particle classification subroutine.

\noindent
Both parts are built using the object oriented technique and C++ language.
The particle identification algorithm is wrapped in FORTRAN closers to be used
in the E781 off-line program.
The package performance was tested in comparison with the likelihood ratio
method using Monte Carlo generated data. Our study has demonstrated the 
excellent ability of the ANN to learn even small details of the detector 
response function.
The ANN solution gives the same performance and behavior as the likelihood
method when using Monte Carlo data with known detector parameters. It
demonstrates that if trained with experimental data the package can provide 
a very good solution to the classification problem of $e^+/e^-$  tracks.

\end{abstract}

\section*{Introduction}
\label{sec:intro}
In this work we have studied the ability of an Artificial Neural Network (ANN) 
based method applied to a Transition Radiation Detector (TRD)
to select $e^+/e^-$ tracks in a high energy fixed target experiment.

In hadroproduction fixed target experiments, such as the E781
experiment at Fermilab~\cite{e781}, the primary 
interaction produces many tracks in a narrow solid angle around the beam 
direction. Most of these particles will not be of leptonic nature.
Nevertheless they can be responsible directly or indirectly (if for instance 
they suffer secondary interactions) for the appearance of ionization clusters 
in the TRD which can fake the $e^+/e^-$ signal under investigation.  

We have used  GE781~\cite{ge781}, the  GEANT~\cite{geant} based Monte Carlo 
simulation of the E781 apparatus in our studies.
The TRD model we are working with is made of 6 blocks, similar to the ones 
described in Ref.~\cite{andreev}, each one composed by a 
radiator (220 foils of CH$_2$) and a Xe-CH$_4$ filled proportional chamber with
the ability to measure the $x$-coordinate of the track.
The detector operates in the cluster counting  mode~\cite{dolgoshein}. 
Ionization clusters were counted only along each track direction to reduce the
influence of the background.

The momentum region of the tracks under investigation is from 10 to 
300 GeV/c, which corresponds to the growing part of the Transition Radiation
yield curve for pions~\cite{dolgoshein}. Above this momentum it is more
difficult to distinguish pions from electrons using this detector.

To achieve appropriate background suppression as well as good signal 
efficiency we have exploited the $\gamma= E/m$ 
(where $E$ is the particle energy and $m$ is its mass) factor 
dependence of the detector response in our algorithm.

In the following we describe and discuss the  maximum likelihood ratio 
technique,  our ANN approach and compare their performance with Monte Carlo 
simulated data. 


\section*{Description of the TRD likelihood method}
\label{sec:likelihood}

This method was proposed for TRD detectors by M. L. Cherry et al.~\cite{MLC} 
and was demonstrated to give better performance than traditional methods in 
Ref.~\cite{AB,KKT,RDA}.

The likelihood function is built in a way to classify particles into two 
categories~: $e^+/e^-$ (type 1) and others, mostly pions, (type 2).
So if a particle of type $i=1,2$ with Lorenz factor $\gamma_i$ generates 
a sample $X$=$\{ x_1, x_2, ..., x_6 \}$ of transition radiation  clusters 
along its track we can define the probability  $P(X|i,\gamma_i)$ 
of this event as 
\begin{equation}   
P(X|i,\gamma_i) = \prod^{6}_{k=1} P(x_k|\gamma_i).
\end{equation}
The probability density functions $P(x_k|\gamma_i)$ were calculated using the 
detector response function.

We can define the likelihood ratio for particle of type $i$ as 
\begin{equation}
 {\cal L}_i (X| \gamma_i) =  \frac{P(X|i,\gamma_i)}{\sum_{j=1}^{2}
 P(X|j,\gamma_j)}, 
\end{equation}
\noindent
which is restricted to the interval $0 \leq {\cal L}_i \leq 1 $,  
and use this ratio as an indicator of the particle type.
For each track one can calculate the above ratio in two possible hypothesis. 
We expect that this ratio will be closer to 1 whenever the hypothesis is 
correct and closer to zero whenever it is wrong. 


\section*{Description of the TRD neural network}
\label{sec:ann}
In order to achieve the maximal performance in the electron 
identification with TRD we have developed an algorithm using a Feed Forward 
Error Back Propagation Artificial Neural Network ~\cite{peterson}.
A similar application was developed in Ref.~\cite{belloti} for 
10 modules of TRD used for cosmic ray lepton identification. 

  As we have 6 modules of TRD 
we will use 6 input nodes with linear response function that will receive the
cluster sum along the track in each TRD block normalized to unity. 
 To explicitly take into account the $\gamma$ dependence of the detector 
response and consequently increase the momentum region in which the algorithm
can provide an efficient classification of tracks, an  extra linear node was 
introduced. This node was fed with a normalized to one $\gamma$ 
factor  calculated in the pion hypothesis, i.e.
\begin{equation}
    \mbox{node 7 activation}  = \frac{E}{m_\pi \gamma_{\mbox{\tiny cut}}},  
\end{equation}
\noindent
where $E$ is the energy of the particle, $m_\pi$ the mass of the pion and 
$\gamma_{\mbox{\tiny cut}}$ was chosen to be 2200, which corresponds to 
pions of about 310 GeV/c.  For $\gamma$ greater than 
$\gamma{\mbox{\tiny cut}}$ the node 7 activation is set equal to one.

\begin{figure}
\epsfverbosetrue
\centering
\epsfysize=5.0truecm
\epsfxsize=7.0truecm
\begin{tabular}{c}
{\epsfbox{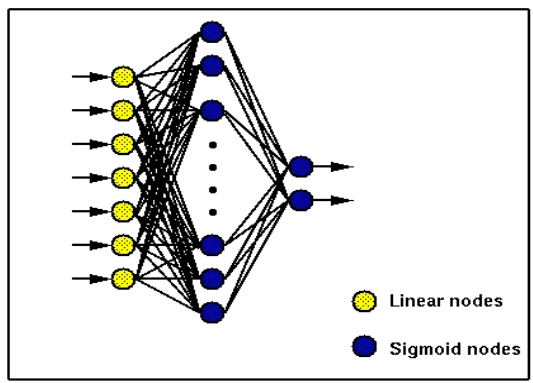}}
\end{tabular}
\caption{The TRD neural network structure.}
\label{fig3}
\end{figure}

We start with 7 input nodes and would like to have a similar classification 
for particles here as in the previous method, that is, two output nodes. So 
according to Kolmogorov theorem~\cite{kolmogorov} we can approximate our 
classification function with 15 nodes in the hidden layer. 
This defines the structure of the network  as 7+15+2 nodes, as 
presented in Fig.~\ref{fig3}.

The first layer of nodes are fully  connected to the second layer of 15 nodes 
 with a sigmoid response function  which is followed by the output layer of 
2 nodes (res 1 and res 2). 
 Each neuron of one layer receives as input the outputs of all neurons from 
the previous layer with weights defined by the synaptic matrix $W$. 
The activation level of the output nodes will provide the track classification.

The sigmoid function we have used is defined as following:

\begin{equation}
 F(x,b)=\frac{1}{1+\exp (-a(x-b))},
\end{equation}

\noindent
where $b$ is the neuron threshold, and $a$ is a gain. It is clear that the 
responses of the two output nodes are bound to the interval $[0,1]$ and 
provide  a similar classification scheme as in the previous method. 

The training process was performed with the standard back-propagation technique 
 ~\cite{peterson} during which  the corrections to the synaptic matrix elements
 $W_{lk}$ were calculated according to the rule :

\begin{equation}
\Delta W_{lk}(i) = - S \frac{\partial E[W]}{\partial W_{lk}}+ M \Delta W_{lk}(i-1),
\end{equation}

\noindent
where $ \Delta W_{lk}(i)$ is the correction to the synaptic matrix elements 
$W_{lk}$ after $i$ steps,  E[W] is the summed square error function, $S$ is 
the learning rate parameter and $M \Delta W_{lk}(i-1) $ is the momentum term 
used to avoid sudden oscillations. We have used in our implementation $S=0.1$
 and $M=0.3$.

 The method consists of two independent parts:
\begin{itemize}
\item the training program;
\item the particle classification subroutine.
\end{itemize}
 Both parts were written in C++ using the Object Oriented (OO) approach. As the
basis for our algorithms we have used modified classes provided by
 Robert Klapper~\cite{sw}.
The training procedure can use either Monte Carlo or experimental data.
The result of the training program is a file describing the ANN structure and
parameters. This file is used then by the particle identification algorithm in
the initialization stage, this permits to dynamically update the ANN 
parameters.
The particle identification algorithm is wrapped in FORTRAN closers to be used
in the E781 off-line program.

To train our network we have used Monte Carlo simulated data with plane
momentum distribution in the whole momentum region. For this the detector
response function was studied and implemented in the Monte Carlo.
It is in principle also possible to use experimental data for this purpose,
 but in this case we need some independent tagging of electrons to prepare 
 pure samples of particles with high statistics and in a wide momentum region.



\section*{Comparison of the methods}
\label{sec:compar}
To compare the performance of these two methods we have used  single tracks 
generated in the whole
momentum region and Monte Carlo simulated hadronic interactions enriched with
leptonic processes. In the last case we have exactly the same momentum
distribution of particles as expected in the experiment.

To choose a suitable efficiency to contamination ratio working point we can 
apply correlated cuts in the feature space of ${\cal L}_e$ (in the $e^+/e^-$
 hypothesis) and ${\cal L}_h$ (in the other particles hypothesis) for the
likelihood method and in the feature space of res 1 and res 2 for the 
artificial neural network. This can be done in the following way:

\begin{equation}
{\cal L}_e>\mbox{cut} ~~~ \mbox{and}~~~{\cal L}_h \leq 1.-\mbox{cut}, 
\end{equation}

\noindent
for the likelihood ratio case and 

\begin{equation}
\mbox{res~1} >\mbox{cut} ~~~ \mbox{and}~~~\mbox{res~2} \leq 1.-\mbox{cut},
\end{equation}

\noindent
for the network, where cut is any real value from zero to one.
Changing the value of the cut we can build plots of the hadronic 
contamination  as a function of  electron detection efficiency for these two 
methods.

\subsubsection*{Single tracks simulation}
\label{sec:mc}

\begin{figure}[t]
\epsfverbosetrue
\centering
\epsfysize=10.0truecm
\epsfxsize=10.0truecm
\begin{tabular}{c}
{\epsfbox{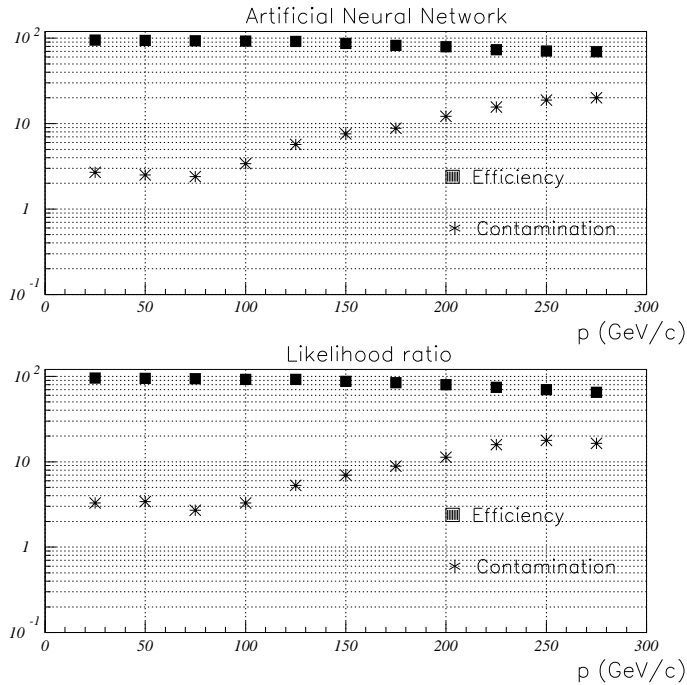}}
\end{tabular}
\caption{Efficiency and contamination for the two algorithms as a function 
of particle momenta.}
\label{fig6}
\end{figure}


We will start our comparison of the methods acceptance and rejection power 
using single tracks generated in a wide momentum range from 10 to 300 GeV/c. 
The momentum region was subdivided in bins of 30 GeV/c, and the corresponding 
parameters were calculated for each bin.
To compare the methods we have selected cuts in such
 a way to have approximately equal efficiencies for both methods in the first 
 momentum bin.


As one can see in Fig.~\ref{fig6} both methods show similar behavior
and the revealed  growth of the contamination and decrease of the efficiency 
corresponds to the degradation with momentum growth of the classification power
 of the detector itself . 


\begin{figure}[hbt]
\epsfverbosetrue
\centering
\epsfysize=8.0truecm
\epsfxsize=10.0truecm
\begin{tabular}{c}
{\epsfbox{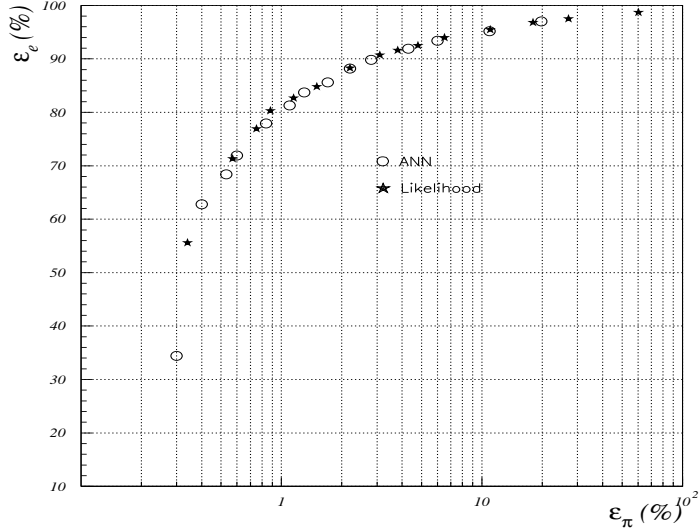}}
\end{tabular}
\caption{Electron identification efficiency ($\epsilon_e$) versus  hadronic 
contamination ($\epsilon_\pi$), taken from simulation of hadroproduction data
 for the two methods under study.}
\label{fig7}
\end{figure}

\subsubsection*{Simulation of the hadronic interaction data}
\label{sec:int}

To compare the two methods in experimental conditions, including hadronic
background, secondary interactions and real momentum distribution we decided 
to use Monte Carlo generated hadronic interactions with complete simulation of 
the detector by the GE781  package. The hadronic background, as well as the 
electromagnetic one, simulated by the package is supposed to be very close 
to the experimental one.

The $e^+/e^-$ efficiency versus the contamination by all other particles for 
such sample is shown in Fig.~\ref{fig7}. One can see that the two methods are
indistinguishable. We really can not give preference to any of these two 
methods from the performance point of view.

\section*{Conclusions}
\label{sec:conclusions}
We have studied the performance of the ANN solution to particle 
classification with a TRD which directly use the momentum information of the 
tracks. 

The ANN solution gives the same performance and behavior as the likelihood
method when using Monte Carlo data with known detector parameters. It
demonstrates that if trained with experimental data the package can provide 
a very good solution if not the best one as the ANN can learn unknown 
properties of the detector and of the experimental conditions that can not 
be implemented in the likelihood method.

\section*{Acknowledgments}
We thank Con\-se\-lho Na\-cio\-nal de De\-sen\-vol\-vi\-men\-to
Cien\-t\'\i \-fi\-co e Tec\-no\-l\'o\-gi\-co~(CNPq) and Fun\-da\c c\~ao de 
Am\-pa\-ro \`a Pes\-qui\-sa do Es\-ta\-do de S\~ao Paulo~(FAPESP) for 
financial support. \\

\end{document}